\documentclass[fleqn,twoside]{article}
\usepackage{espcrc2}

\usepackage{epsfig}
\usepackage[figuresright]{rotating}

\newcommand{\AmS}{{\protect\the\textfont2
  A\kern-.1667em\lower.5ex\hbox{M}\kern-.125emS}}

\title{A Cannonball Model of Cosmic Rays}

\author{A. De R\'ujula\address[MCSD]{Theory Division, CERN;
1211 Geneva 23, Switzerland\\
Physics Department, Boston University, USA}}
       
\begin{document}

\begin{abstract}
I outline a {\it Cannon Ball} model of Cosmic Rays in which their distribution in the Galaxy, 
their total ``luminosity", the broken power-law spectra with 
their observed slopes, the position of the knee(s) and ankle(s), and the alleged 
variations of composition with energy are all explained in terms of 
simple and ``standard"  physics. 
%The model is  lacking a satisfactory 
%theoretical understanding of the ``cannon" that emits the cannonballs in 
%catastrophic episodes of accretion onto a compact object.

\vspace{1pc}
\end{abstract}

\maketitle

\section{Credits} 	%) A SECTION HEADING

The {\it Cannonball} (CB) model is a unified model of high-energy
astrophysics, in which the gamma background radiation, cluster ``cooling
flows", gamma-ray bursts, X-ray flashes and cosmic-ray electrons and
nuclei of all energies ---share a common origin. The mechanism underlying all
these phenomena is the emission of relativistic ``cannonballs" by ordinary
supernovae, analogous to the observed ejection of plasmoids by 
quasars and microquasars.
It is not unusual in talks to start with the credits, as in a film.
Many of the ideas I shall exploit have a long pedigree: Gamma-Ray
Bursts (GRBs) are the main (injection) process for Cosmic Rays \cite{DKNR} (CRs); 
GRBs are the main CR (production and acceleration) mechanism \cite{DP},
and are induced by narrow jets emitted by accreting compact stellar
objects \cite{SD}; their $\gamma$-rays being low-energy photons boosted to higher
energies by inverse Compton scattering \cite{SD} (ICS). The concrete realization 
of these ideas in the ``CannonBall" (CB) model is more recent
and covers GRBs \cite{GRB1,GRB2}, X-Ray Flashes \cite{XRF} (XRFs), their respective
afterglows \cite{AGoptical,AGradio}, the Gamma ``Background" Radiation \cite{GBR}, the 
CR luminosity of our Galaxy \cite{CRL}, 
the ``Cooling Flows" of galaxy clusters \cite{CF},
and the properties of CRs \cite{DP,CRArnon,Yo}. 

\section{Jets in Astrophysics}

\begin{figure}[]
%\vskip -3.4cm
\centering
\vbox{%\hskip -9.5mm
 \epsfig{file=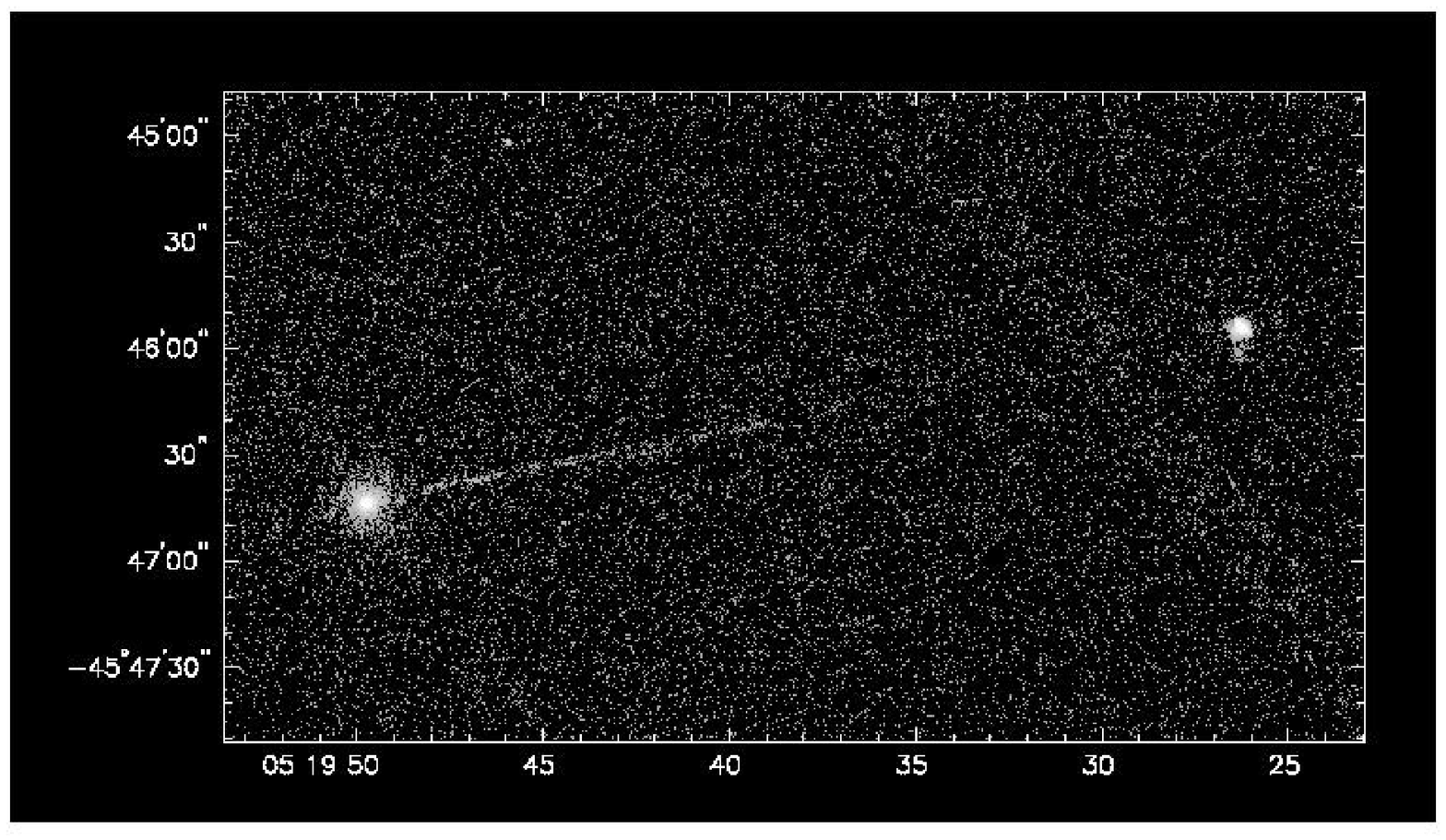,height=4cm,width=7cm}
% \includegraphycs*{file=PictorAjet_mod.eps,height=4.5cm,width=7.1cm}
}
\vbox{%\hskip -.5mm
 \epsfig{file=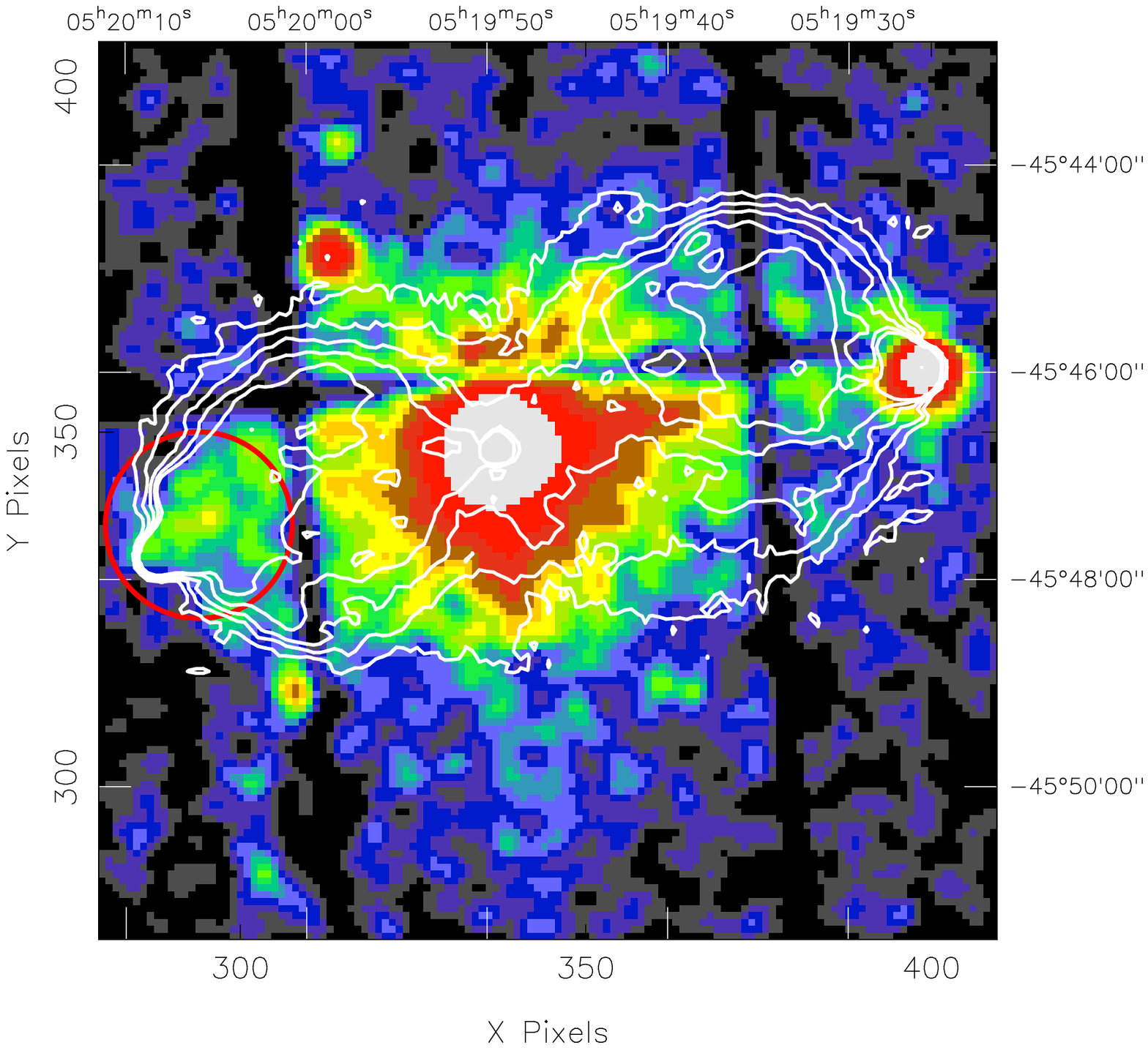,height=8cm,width=7.cm,angle=-90}
}
\vspace*{-15pt}
 \caption{Above: X-ray image of the  galaxy 
Pictor A: a non-expanding jet extends
across 360000 light years towards a hot spot at least 800000
light years away from where the jet originates. Below: 
XMM/p-n image of Pictor A in the 0.2--12 keV energy interval,
centred at the position of the leftmost spot in the upper panel,
superimposed on the radio
contours of a 1.4 GHz radio VLA map.}
 \label{Pictor}
\end{figure}

A look at the sky, or a more modest one at the web, results in the
realization that jets are emitted by many astrophysical systems
(stars, quasars, microquasars...). One  impressive case \cite{Wilson} 
is that of the quasar Pictor A, shown in Fig.~\ref{Pictor}. {\it Somehow},
the active galactic nucleus of this object is discontinuously 
spitting {\it something} that does not
appear to expand sideways before it stops and blows up, having by then
travelled for a distance of several times the visible radius of a
galaxy. Many such 
systems have been observed. They are very relativistic: the Lorentz factors (LFs)
$\gamma\equiv E/(mc^2)$ of their ejecta are typically of 
${\cal{O}}(10)$.
The mechanism responsible for these mighty ejections ---suspected
to be due to episodes of violent accretion into a very
massive black hole--- is not understood.

In our galaxy there are ``micro-quasars'', in which the central black
hole has only a few times the mass of the Sun. The first  
example \cite{Felix} was the $\gamma$-ray source  GRS 1915+105.
Aperiodically, about once a month, it emits two
opposite {\it cannonballs}, travelling at $v\sim 0.92\, c$.
As the event takes place, 
%and as illustrated in Fig.~\ref{fig:5}, 
the X-ray 
emission ---attributed to an unstable accretion disk--- temporarily decreases.
How part of the accreating material ends up ejected along the system's axis
is not understood. The process reminds one of the blobs emitted
upwards as the water closes into the ``hole'' made by a stone
dropped onto its surface. For quasars and $\mu$-quasars, 
it is only the relativistic, general-relativistic
magneto-hydro-dynamic details that remain to be filled in!
Atomic lines from many elements have been observed \cite{Kotani} in
the CBs of $\mu$-quasar SS 433. Thus, at least in this case, the
ejecta are made of ordinary matter, and not of some fancier substance
such as $e^+e^-$ pairs.

\section{The Cannonball Model}

\begin{figure}
\centering
 \epsfig{file=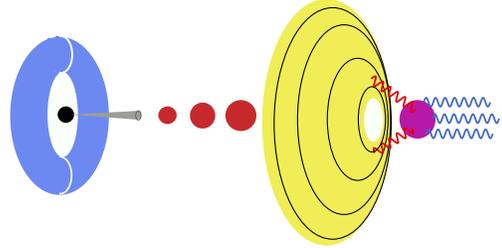,width=0.44\linewidth,angle=-90}
\vspace*{-20pt}
\caption{The CB model
of long-duration GRBs \cite{GRB1}. A core-collapse SN results in
a compact object and a fast-rotating torus of non-ejected
fallen-back material. Matter (not shown) abruptly accreting
into the central object produces
a narrowly-collimated beam of CBs, of which only some of
the ``northern'' ones are depicted. As these CBs move through
the ``ambient light'' surrounding the star, they Compton up-scatter
its photons to GRB energies \cite{GRB2}.}
\label{figCB}
\end{figure}

The ``cannon'' of the CB model is analogous to the ones
responsible for the ejecta of quasars and microquasars.
{\it Long-duration} GRBs, for instance, are produced in
{\it ordinary core-collapse} supernovae (SNe) by jets of CBs, made of {\it
ordinary-matter plasma}, and travelling with high Lorentz factors (LFs),
$\gamma\sim{\cal{O}}(10^3)$. An accretion torus is hypothesized 
to be produced around
the newly-born compact object, either by stellar material originally
close to the surface of the imploding core and left behind by the
explosion-generating outgoing shock, or by more distant stellar matter
falling back after its passage \cite{ADR,GRB1}. A CB is emitted, as
observed in microquasars \cite{Felix}, when part of the accretion disk
falls abruptly onto the compact object, see Fig.~\ref{figCB}.

\begin{figure}
\centering
 \epsfig{file=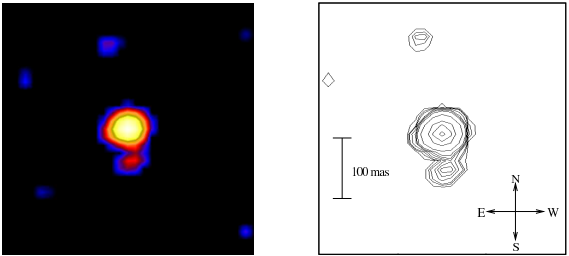, width=1\linewidth}
%\hskip .5truecm
\vspace*{-18pt}
\caption{The two CBs emitted by SN1987A in opposite axial
directions. The northern and southern
bright spots are compatible with  CBs emitted at
the time of the SN explosion and travelling at a velocity equal,
within errors, to $c$. One of the {\it apparent} velocities is superluminal.
The corresponding GRBs were not pointing in our direction, which
may have been a blessing.}
\label{figCostas}
\end{figure}

{\it Do supernovae emit cannonballs?} Up to last year, there was only one
case in which the data was good enough to tell: SN1987A, the core-collapse
SN in the LMC, whose neutrino emission was detected. Speckle interferometry
measurements made 30 and 38 days after the explosion \cite{NP} 
did show two relativistic CBs (one of them ``superluminal''), 
emitted in opposite directions, as shown in Fig.~\ref{figCostas}.

\section{GRB Afterglows and Cosmic Rays}

A freshly jetted CB is assumed to be expanding at a speed comparable to that
of sound in a relativistic plasma ($c/\sqrt{3}$). Its typical baryon number
 is that of half of Mercury, $N_{_{CB}}\!\sim\! 10^{50}$,
its start-up LF is $\gamma_0\!\sim\! 10^3$ (both ascertained from 
the properties of afterglows (AGs), and of the GRB's $\gamma$ rays). In their voyage,
CBs intercept the electrons and nuclei of the interstellar medium
(ISM), previously ionized by the GRB's $\gamma$ rays.
In seconds of (highly Doppler-foreshortened) observer's time, such an 
expanding CB becomes ``collisionless'', that is, its radius becomes bigger than 
a typical nucleus-nucleus interaction length. But it still interacts with the
charged ISM particles it encounters, for it contains a  magnetic field\footnote{
Numerical analysis of the merging of two plasmas at a high relative
$\gamma$, based on following each particle's individual trajectories
as governed by the Lorentz force and Maxwell's equations, demonstrate
the generation of such turbulent magnetic fields, as well as the ``Fermi'' 
acceleration of particles, in the total {\it absence} of shocks \cite{Fred},
 to a power law spectrum: $dN/dE\approx E^{-\beta_s}$,
 with $\beta_s\sim 2.2$.}. 

If the nuclei entering a CB are magnetically ``scrambled''
and are reemitted isotropically in the CB's rest system, a radial loss of
momentum results. The rate of such a loss corresponds to an inwards radial
force on the CB. In our analysis of GRB AGs, we assumed
%\footnote{These assumptions are unnecessary to the CB model {\it of CRs.}}: 
that this force counteracts the expansion,
and that when the radius stabilizes, the inwards pressure is in equilibrium
with the pressure of the CB's magnetic field. 
This results in values for the asymptotic CB radius
($R_{_{CB}}\!\sim\! 10^{14}$ cm for typical parameters) and its time-dependent
magnetic-field strength \cite{AGradio}:
\begin{equation}
B_{_{CB}}[\gamma(t)]=3\;{\rm Gauss}\;{\gamma(t)\over 10^3}\;
\left({n_p\over 10^{-3}\;{\rm cm}^{-3}}\right)^{1/2}\; ,
\label{B}
\end{equation}
where $n_p$ is the ISM number density, normalized to a value characteristic
of the ``superbubble'' domains in which SNe and GRBs are born. Our 
assumptions are no doubt na\"ive, but they are to be
judged in light of two facts: 1) The very simple ensuing 
analysis \cite{AGoptical,AGradio}
of the elaborate time and frequency dependence of AGs,  
dominated by electron synchrotron radiation of in the field of Eq.~(\ref{B});
2) The CBs emitted by certain objects appear not 
to expand significantly, as in the example in the upper part of Fig.~\ref{Pictor}.

As a CB pierces through the ISM, its LF, $\gamma(t)$, continuously
diminishes, as its energy is dominantly transferred to scattered ISM nuclei,
and subdominantly to scattered electrons and synchrotron photons. All
these reemitted particles, in the rest system of the host galaxy, are forward-peaked
in distributions of characteristic opening angle $1/\gamma(t)$. In the lower 
Fig.~\ref{Pictor} the two jets of Pictor A are shown, with contour plots corresponding
to radio-intensity levels \cite{Grandi}. 
We interpret this radio signal as the synchrotron radiation of 
the CB-generated {\it Cosmic-Ray electrons}  in the ambient magnetic fields.
The CBs of Pictor A
must also be scattering the intercepted  nuclei, and converting them into
{\it Cosmic-Ray Nuclei}.

The range of a CB is governed by the rate at which it loses momentum
by encountering ISM particles and catapulting them into CRs.
The initial LF, $\gamma_0\!\sim\! 10^3$,
is typically halved in a fraction of a kpc, while a CB becomes 
non-relativistic only at distances of 10's or even 100's of kpc,
well into a galaxy's halo or beyond. The CRs of the CB model are
deposited along the long {\it line} of flight of CBs, in contradistinction
with those of the standard models, in which the CRs are generated
by SN shocks,
%\cite{Shkl}, 
at the {\it ``points''} where they occur in ``active''
regions of stellar birth and death. In the CB model no reacceleration 
 far from the CR birth-sites need be invoked to accommodate the 
data. This is most relevant for electrons, which lose energy fast,
and locally  \cite{GBR}.

\section{The ``GBR" and the CR electrons}

\begin{figure}[]
\centering
\vbox{
 \epsfig{file=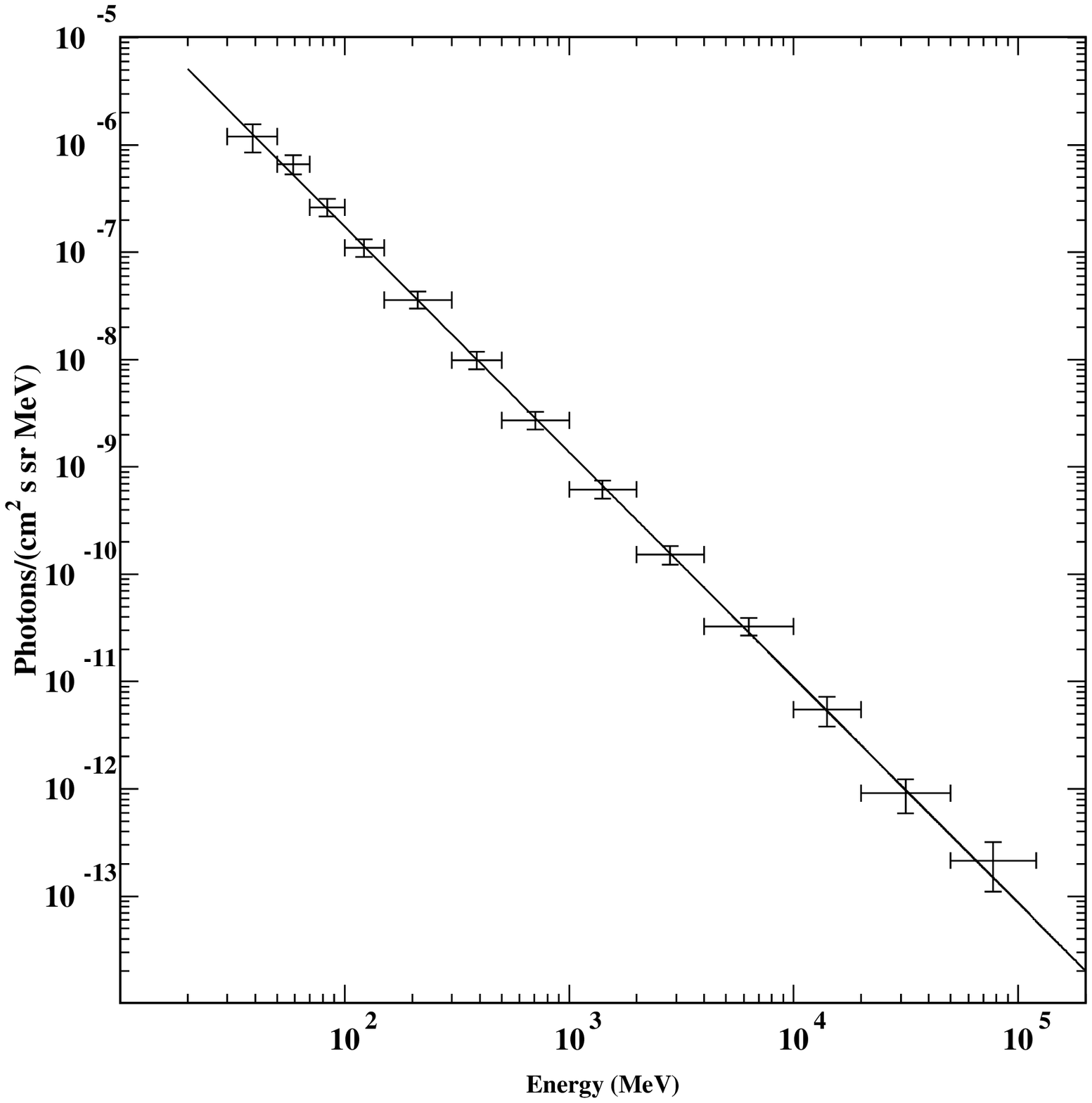,width=0.48\linewidth}
%\hspace*{0.5mm}
%\epsfig{file=                  gbr0.eps,width=6cm}
 \epsfig{file=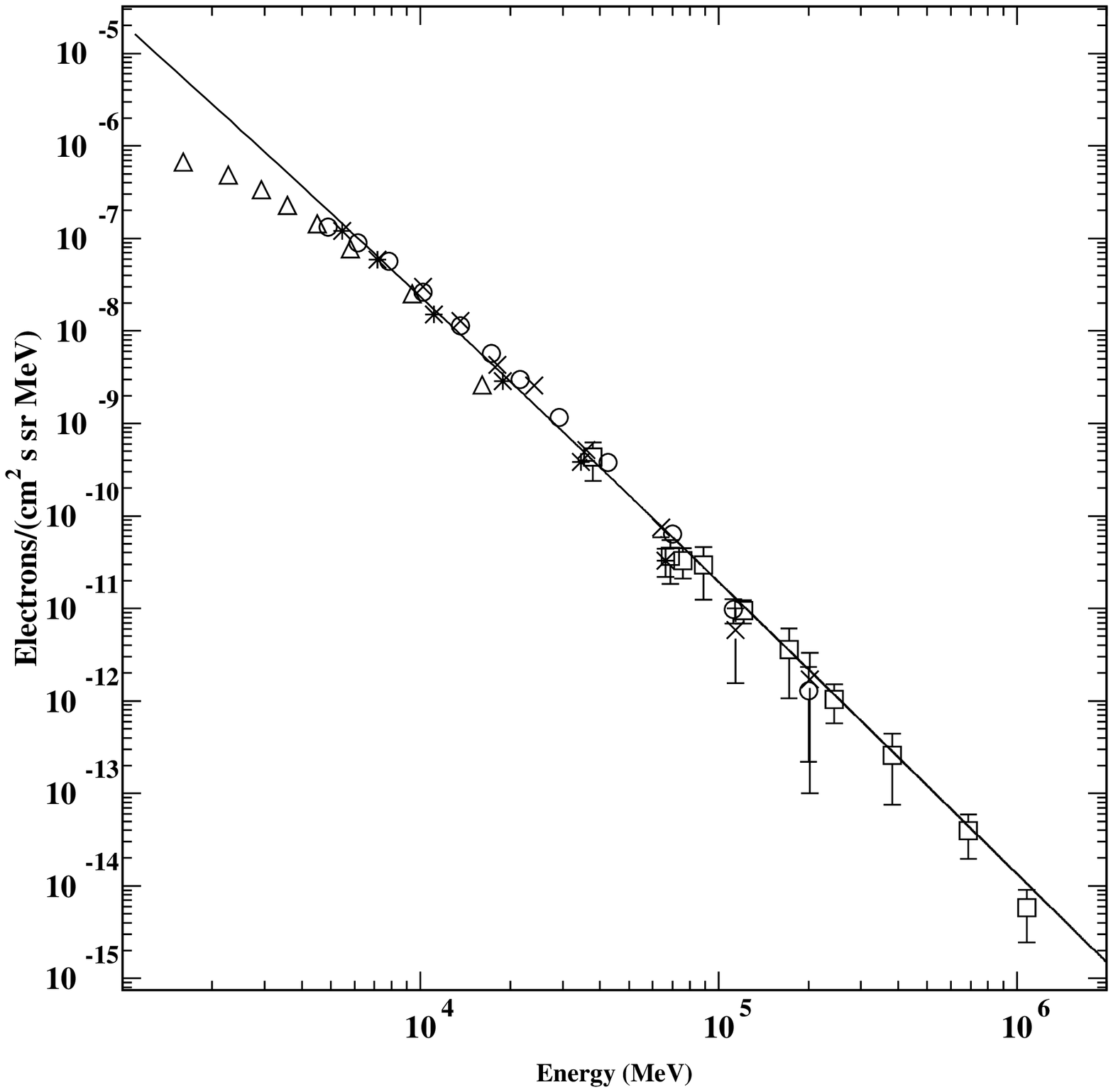,width=0.48\linewidth}
}
\vspace*{-14pt}
\caption{Left: Comparison between the spectrum  of
the GBR, measured by EGRET,
and the prediction (the line) for ICS of starlight and the CMB by CR
electrons.  Right:
The primary CR electron spectrum. The slope is the prediction, the
magnitude is normalized to the data.}
 \label{eGBR}
\end{figure}

The existence of an isotropic, diffuse gamma background radiation (GBR,
confusingly similar to GRB)
was first suggested by data from the SAS 2 satellite \cite{TF}.
The EGRET/CRGO instrument 
confirmed it:  {\it ``by removal of point sources and of the
galactic-disk and galactic-centre emission, and after an extrapolation
to zero local column density'',}
a uniformly distributed GBR was found, of alleged
extragalactic origin \cite{Sree}. Above an energy of  
$\sim\!10$ MeV, this radiation  has a featureless spectrum, shown in Fig.~\ref{eGBR},
which is very well described by a simple power-law form, 
$dF/dE\propto E^{-\beta_{_{\rm GBR}}}$, with 
$\beta_{_{\rm GBR}}\approx 2.10\pm0.03$.

%
%\begin{figure}
%%\plotone{gbrfig1.eps}
%\begin{center}
%\vspace*{-1.6cm}
%\hspace*{-1cm}
%\epsfig{file=gbrfig1.eps,width=10cm}
%\caption{Comparison between the spectrum  of
%the GBR, measured by EGRET (Sreekumar et al.~1998),
%and the prediction for ICS of starlight and the CMB by CR
%electrons. The slope is our central prediction, the normalization
%is the one obtained for ${\rm h_e}= 20$ kpc, ${\rm \rho_e}= 35$ kpc.}
%\vspace*{-0.5cm}
%\end{center}
%\end{figure}
%\begin{figure}
%\begin{center}
%\vspace*{-1.6cm}
%\hspace*{-1cm}
%\epsfig{file=gbr0.eps,width=10cm}
%%\vspace*{-14.6cm}
%\caption{The primary cosmic-ray electron spectrum
%(Evenson \&
%Meyers 1984; Golden et al.~1994; Ferrando et al.~1996) as
%measured by Prince 1979 [crosses]; Nishimura et al.~1980 [squares];
%Tang 1984 [circles]; Golden et al.~1984 [triangles];  Barwick et
%al.~1998 [stars]. The slope is the prediction, the
%magnitude is normalized to the data.}
%\vspace*{-0.5cm}
%\end{center}
%\end{figure}

There is no consensus on what the origin of the GBR is.
The proposed candidate sources range from the
rather conventional (e.g.~active galaxies \cite{Big}) to the decisively 
speculative (e.g.~primordial black hole evaporation \cite{PH}).
A ``cosmological'' origin is the most noble putative ancestry, but
though the GBR index $\beta_{_{\rm GBR}}$ is uncannily direction-independent,
the EGRET GBR flux in directions above the galactic disk and centre
shows significant anisotropies, correlated with our position relative
to the centre of the Galaxy \cite{DDA}. How does the GBR relate to CRs?

Below a few GeV, the local spectrum of CRs is affected by the solar
wind and the Earth's magnetic field, its modelling is elaborate.
Above $\sim\! 5$ GeV, the spectrum of CR electrons, shown in Fig.~\ref{eGBR},
is well fit by a power law \cite{Ferr}: 
$dF/dE\propto E^{-\beta_e}$, with $\beta_e\approx 3.2\pm0.1$.
The nuclear CR spectrum,
above $\sim\! 10$ GeV and up to the ``knee'' at $\sim 3\times 10^6$ GeV,
 is also a single power law: 
$dF/dE\propto E^{-\beta_p}$, with $\beta_p\approx 2.70\pm0.05$.
%($\sim\,$96\% of these nuclei are protons, at a fixed energy {\it per nucleon}).

As discussed in detail in Section 7.1, CBs accelerate the ISM
electrons and nuclei they encounter in their path as a ``magnetic-racket''
would, imparting to all species the same distribution of LFs $\gamma$.
This means that the {\bf source} spectra of (relativistic) nuclei and electrons 
have the same energy dependence: $dF/dE\propto E^{-\beta_s}$, with
a species-independent $\beta_s$. The observed spectra are not the
source spectra. The nuclear flux is modulated by the energy dependence
of the CR confinement-time, $\tau_{\rm conf}$, in the magnetized disk and halo of the
galaxy, affecting the different species in the same way, {\it at fixed} $E/Z$,
with $Z$ the nuclear charge. Confinement effects are not  understood, but
observations of astrophysical plasmas and of CR  
abundances as functions of energy suggest \cite{Swo}:
\begin{equation}
\tau_{\rm conf}\propto \left({Z/ E}\right)^c\; ,
\label{c}
\end{equation}
with $c\sim 0.5\pm 0.1$ at the low energies at which the CR composition
is well measured. This means that $\beta_s=\beta_p-c\sim 2.2$, as in 
footnote 1.

Above a few GeV,  the electron spectrum is dominantly
modulated by ICS on starlight and on the microwave background radiation, 
the corresponding electron ``cooling'' time being shorter than their confinement 
time. For an equilibrium situation between electron CR generation and ICS
cooling, this implies that $\beta_e=\beta_s+1\sim 3.2$, a prediction \cite{GBR}
in perfect agreement
with observation, as in Fig.~\ref{eGBR}. In the CB model, the Compton upscattered
photons {\bf are} the GBR, and their spectrum  is a power law with a predicted \cite{GBR}
 index  $\beta_{_{\rm GBR}}\!=\!(\beta_e+1)/2\!\sim\! 2.1$, also in agreement
with the data, as in Fig.~\ref{eGBR}. Cannonballs deposit CRs along their 
linear trajectories, which
extend well beyond the Galaxy's disk onto the halo and beyond. The observed
non-uniform (i.e.~non-cosmological) distribution of GRB flux in intensity, latitude and 
longitude is well reproduced \cite{GBR} for an ellipsoidal CR halo of 
--within very large errors-- characteristic height 
$\sim\!20$ kpc, and radius $\sim \! 35$ kpc.

\section{The CR Luminosity of the Galaxy}

If the CRs are chiefly Galactic in origin, their accelerators must
compensate for the escape of CRs from the Galaxy to sustain the
observed CR intensity: it is known from meteorite records that
the CR flux has been fairly steady for the past few giga-years \cite{Lon}.
The conventional estimate of the CR luminosity of the Milky Way is
%Milky Way's luminosity in CRs must therefore satisfy:
%\begin{eqnarray}
%L_{CR}&\approx& L_p ={4\pi \over c} \int {1\over \tau_{\rm conf}}
%\,E\,{dF_p\over dE}\,dE\,dV
%\nonumber\\
%&\sim& {4\pi \over c}
%\int \bar\rho \,dV\int {1\over X}\,E\, {dF_p\over dE}\,dE ,
%\label{CRlum}
%\end{eqnarray}
%where the last result is the standard estimate, thus obtained:
%The mean column density $X$ traversed by CRs before they reach the Earth
%can be extracted\cite{Swo} from the observed ratios of primary CRs to 
%secondaries (products of spallation).
%% : $X\approx 6.9\,[E({\rm GeV})/(20\, Z)]^{-0.5}$ g cm$^{-2}$.
%With use of ${X=\int \rho\, dx\sim\bar\rho\, c\, \tau_{\rm conf}}$ one can
%extract the product of ${\tau_{\rm conf}(E)}$ and a path-averaged
%density $\bar\rho$. If the local values of $X$ and ${dF_p/dE}$
%are representative of the Galactic values dominating the first integral in
%Eq.~(\ref{CRlum}), the final result follows.
%Assume the path-averaged $\bar\rho$ to be close to the average
%density $\rho$ of neutral and ionized gas in the Galaxy, so that
%${ \int\bar\rho\, dV}$ is the total mass of Galactic gas,
%estimated\cite{Lon} to be
%${ 5\times 10^9\, M_\odot}$.
%The numerical result is
\cite{Dru}:
\begin{equation}
L_{CR} \sim 1.5\times 10^{41}~{\rm erg~s^{-1}}\, .
\label{them}
\end{equation}

In the CB model $L_{CR}$ can be estimated from the electron CR density
involved in its successful description of the GBR, if the local
observed ratio of proton to electron fluxes is representative of the
Galactic average. It can also be estimated from the rate of Galactic SNe and
the typical energy in their jets of CBs. The results of these  estimates 
agree \cite{Lum},
but they are over one order of magnitude larger than Eq.~(\ref{them}). This is
not a contradiction, for the CB-model effective volume of confined CRs
is much bigger than in the standard picture, wherein CRs are confined
to the Galactic disk. 
The CB-model value of the CR confinement time 
 is also one order of magnitude larger than the
standard result, based on the ratios of stable to unstable isotopes \cite{Lum}. 
This alterity is understood \cite{Plaga}: the stable
CRs spend much of their  time in the Galaxy's halo,
which in the CB-model is magnetized by the flux of CRs that
the CBs deposit in it.

\section{The Cosmic Ray Spectra}

\begin{figure}
\centering
 \epsfig{file=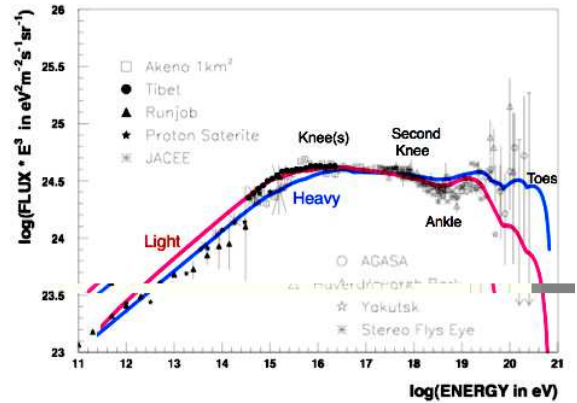,width=\linewidth}
\vspace*{-25pt}
\caption{The ``all-particle'' CR spectrum. The ``light'' and ``heavy''
 lines are predictions for the two CR  abundances discussed in Section 7.9.
 Only one parameter was adjusted, see Section 7.5.}
\label{allpart}
\end{figure}

It is customary to ``renormalize'' the energy calibration of
different experiments to make flux measurements look in better
agreement; and to present the data as the flux times a power of energy,
to emphasize the ``features'' of the spectrum and its
changes of power ``index''. This is done in Fig.~\ref{allpart}
for the {\it ``all-particle''} spectrum \cite{CRspec}, showing the {\it ``knee''}
at (2 to 3) 
$\times \;10^{15}$ eV, the {\it ``second knee''} at $\sim 5 \times 10^{17}$
eV and the {\it ``ankle''} at $\sim 3\times 10^{18}$ eV. The purpose of
this section is to outline how these features, and the changes 
of CR composition with energy, are simple consequences of the
CB model of CR production.

\subsection{``Collisionless magnetic rackets''}

In an elastic collision of a relativistic CB of LF $\gamma$
with (much lighter) ISM electrons or ions at rest,  
the light recoiling particles (of mass $M$) have an energy spectrum
extending up to $E=2\,\gamma^2\,M\,c^2$.
This is a {\it magnetic-racket accelerator} of gorgeous efficiency:
the ISM particles reach up to $\Gamma=2\,\gamma^2$.
Single elastic scattering of target particles at rest
is not the whole story, for CBs may collide with previously-accelerated
CRs. Also, as in footnote 1, CBs may internally ``Fermi''-accelerate particles, 
before reemitting them. The extreme in which the first process is dominant
has been studied by Dar \cite{CRArnon}. Here, the opposite extreme \cite{Yo} is
discussed. A bit coincidentally, the two extremes lead to very
similar results. 

In our study of GRB AGs, we assumed that the AG is dominated by
electron synchrotron radiation in the magnetic field of Eq.~(\ref{B}).
We also used the simplifying assumption that half of the electrons 
within a CB are unaccelerated, while the other half are accelerated,
in the CB's rest system, to a  source spectrum 
$dN/d\gamma_e\propto\gamma_e^{-\beta_s}\,\Theta(\gamma_e-\gamma)$, with 
$\beta_s=2.2$ (as in footnote 1), and
$\gamma=\gamma(t)$ the LF or the incoming electrons: the one of the 
CB in the SN rest system.
This leds to the prediction of a wide-band AG spectrum in excellent
agreement with observations \cite{AGradio}. Here, I make similar assumptions:
the LF distributions of the ISM nuclei intercepted, magnetically
deflected, partially accelerated and reemitted by a CB are {\it identical}
to those of electrons (but for the effect of electron cooling by synchrotron radiation).
The only other difference is that we shall discuss
 nuclei of up to very high energies, for which the ``Larmor'' limit
---on the maximum possible acceleration within a CB---  plays a role.

\subsection{Elastic scattering: the ``knees''}

Let $m$ be the proton mass and $\sim\!m\,A$ that of
a nucleus of atomic weight $A$. The ISM nuclei recoiling from an elastic 
scattering with a CB of LF $\gamma$ have energies in the range 
$m\,A\le E_A\le 2\,m\,A\,\gamma^2$. The initial LFs of CBs,
extracted from the analysis of their AGs \cite{AGoptical,AGradio} and/or 
``peak energies'' \cite{GRB1,GRB2} 
peak at $\gamma_0\sim 10^3$ and have a narrow distribution extending
 up to  $\gamma_0\sim 1.5\times 10^3$. Thus, the spectrum of nuclei elastically
 scattered by CBs should end at an energy:
 \begin{equation}
 E[{\rm knee}]\sim (2\;{\rm to}\;4)\,10^6\,A\;{\rm GeV.}
 \label{Eknee}
 \end{equation}
 We shall see anon that $E[\rm knee]$ is also the position at which the
spectrum of inelastically scattered nuclei changes its slope.

\begin{figure}
\centering
 \epsfig{file=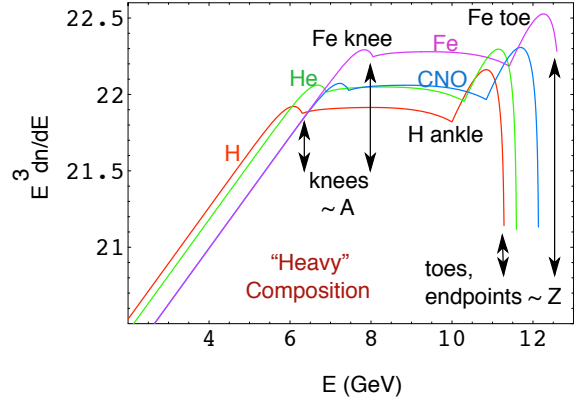, width=1\linewidth}
\vspace*{-29pt}
\caption{Log-log plot of the predicted $E^3\,dN/dE$
spectra of H, He, the CNO group and the Fe group, for the ``heavy''
CR relative abundances of Section 7.9. The relative abundances
become more ``metallic'' at the knees and again above the ankle.
Only one parameter was adjusted.}
\label{ffff}
\end{figure}

The individual spectra of abundant CR elements and groups  are
shown in Fig.~\ref{ffff}.
Preliminary CR composition data from  KASCADE  \cite{KASKA} indicate
that there is a change of slope of the individual spectra at the values
predicted in Eq.~(\ref{Eknee}), but the data are not yet good enough to 
establish the predicted linear $A$-dependence, or to distinguish it from
a putative $Z$-dependence.

\subsection{``Accelerated scattering'': the ``toes''}

The spectrum of Fermi-accelerated particles within a CB cannot extend
beyond the $Z$-dependent energy at which their Larmor radius 
in the CB-field of Eq.~(\ref{B}) is larger than the CB's radius. For
typical parameters:
\begin{equation}
E[{\rm Larmor}]\simeq 9\times 10^{16}\;Z\;{\rm eV}\;
{B_{_{CB}}[\gamma_0]\over 3\;{\rm G}}\;
{R_{_{CB}}\over 10^{14}\;{\rm cm}}\; .
\label{Larmor}
\end{equation}
The ISM nuclei exiting a CB after having being accelerated within it have
energies extending up to $E[{\rm toe}]=2\,\gamma_0\,E[{\rm Larmor}]$, that is:
\begin{equation}
E[{\rm toe}]\sim(2\;{\rm to}\;6)\,10^{11}\,Z\;{\rm GeV},
\label{toe}
\end{equation}
which is the maximum energy to which the CB mechanism we have
discussed is capable of accelerating CRs. Notice that the predicted ``toes'' at
the spectral end scale as $Z$, not like $A$, as the ``knees'' do, as illustrated
in Fig.~\ref{ffff}.

The energy $E[\rm toe]$ for Fe nuclei is comparable to the maximum observed
ones in CRs. There is some evidence \cite{HEcompo} for changes 
of composition above the ankle,
compatible with those implied by Fig.~\ref{ffff}. But
the extraction of relative CR abundances at very high energies is a difficult task.

\subsection{The deceleration of CBs in the ISM}

Consider a CB of initial mass $M_0$, traveling through the ISM at an instantaneous
LF $\gamma$. Let $a$ be the ratio between the average energy of a nucleus exiting
a CB in its rest system and the energy at which the nucleus entered, so that
$\langle \gamma_{out}\rangle\equiv a \, \gamma$. For elastic scattering, $a=1$;,
for nuclei fagocitated by the CB, $a=0$; and for those Fermi-accelerated within
the CB, $a\!>\! 1$. Let $\bar a$ be the mean value in the average over these
processes, and $\bar A$ the mean weight in the ISM density distribution
$dn_{_A}$. For $\gamma^2\gg 1$, energy-momentum conservation implies
 a CB's deceleration law:
\begin{equation}
{d\gamma\over\gamma^k}\simeq-{m\over M_0}\,{\gamma_0^{\bar a-1}}\; 
{\bar A}\;dn_{_A}\; ,\;\;\;\;\;\;
k\equiv3-\bar a\, .
\label{decelerate}
\end{equation}
To compute the spectrum of the CRs produced by a CB in its voyage through
the ISM we have to let the CB decelerate from $\gamma=\gamma_0$
to $\gamma\sim 1$, tantamount to integrating the CR spectra at
local values of $\gamma$ with a weight factor 
$dn_{_A}\propto {d\gamma/\gamma^k}$.

The value of $k$ in Eq.~(\ref{decelerate}) cannot be ascertained with confidence.
One reason is the averaging over the quoted processes. Three other reasons are:
1) The nucleus-CB elastic scatterings may not be isotropic. If the cross-section
is modeled as a power-law in momentum transfer,
$d\sigma\propto (1+\beta\cos\alpha)^{-a1}$, one obtains an approximate
``effective'' deceleration law $d\gamma/\gamma^{k1}$, with $k1\!>\!k$. 2)
The reemission of accelerated nuclei may be delayed, so that they exit the CB 
at $\gamma_{exit}\!<\!\gamma$. If the $\gamma_{exit}$ distribution is
modeled as a power law, $\gamma_{exit}^{-a2}$, 
one obtains again an approximate
``effective'' deceleration law $d\gamma/\gamma^{k2}$, with $k2\!>\!k$. 3)
Slow CBs would be unobservable in GRBs or their AGs, for the fluences are
biased towards large $\gamma$. Microquasars may also contribute  
low-$\gamma_0$ CR-generating CBs. If the $\gamma_0$ distribution is
modeled as a power law, once again the ``effective'' value of $k$ is increased.
All in all, we may expect $k\sim3$, but we cannot predetermine its value.

\subsection{Spectrum of elastically-scattered CRs}

Let the elastically scattered CRs, exiting a CB in its rest system with the 
decelerating instantaneous value of $\gamma$, be isotropically emitted,
with a constant $d\sigma/d\cos\alpha$: we have seen that
a reasonable non-isotropy only leads to an increase of $k$. Boosted by
the CB's motion, the instantaneous CR spectrum in the SN rest system is:
\begin{eqnarray}
{dN\over d\gamma_{_A}}&\propto&\int_{-1}^{1}{d\cos\alpha\over 2}\;
\delta[\gamma_{_A}-\gamma\,\gamma\,(1+\cos\alpha)]\nonumber\\
&=&{1\over 2\,\gamma^2}\;\Theta[2\gamma^2-\gamma_{_A}].
\label{elastic1}
\end{eqnarray}

To obtain the total ``elastic'' CR spectrum, integrate over the
CB's trajectory:
\begin{eqnarray}
&&\!\!\!\!\!\!\!\!\int_{\rm traj}dn_{_A}\,{dN\over d\gamma_{A}}\propto
\int_1^{\gamma_0}{d\gamma\over\gamma^k}\;{dN\over d\gamma_{_A}}\propto
\label{traj}\\
&&\!\!\!\!\!\!\!\!
 \left[{1\over\gamma_{_A}}\right]^{k+1\over 2}
\left[1-\left({\gamma_{_A}\over 2\,\gamma_0^2}\right)^{k+1\over 2}\right]
\Theta[2\gamma_0^2-\gamma_{_A}].
\label{elastic2}
\end{eqnarray}
This elastic-scattering contribution extends up to
$\gamma_{_A}=2\gamma_0^2$, as announced in Section 7.2.
For energies below these knees,  the Galaxy confines CRs
so that the result of Eq.~(\ref{elastic2}) is to be modified by
the multiplicative factor $\propto 1/ \gamma_{_A}^c$ of Eq.~(\ref{c}).
The observed slope $\beta\sim 2.7$ of the CR spectra below the
knees is reproduced for $c+(k+1)/2=\beta$. In practice, this combination
of parameters is the {\it only quantity} chosen by hand in
predicting the all-particle and individual CR spectra.

\subsection{The spectrum of CB-accelerated CRs}

The spectrum of nuclei accelerated within a CB is 
``flavour-blind'' in the variable $\gamma_{_A}$, and of the form
$dN/d\gamma_{_A}\propto\gamma_{_A}^{-\beta_s}\,
\Theta(\gamma_{_A}-\gamma)\,\Theta(b\,\gamma-\gamma_{_A})$, 
with $\beta_s=2.2$, and
$\gamma=\gamma(t)$. The second $\Theta$ function is the Larmor cutoff,
for typical parameters $b\sim 10^5$. Boosted to the SN rest system, 
the instantaneous CR spectrum is:
\begin{eqnarray}
&&\int_\gamma^{b\gamma}{d\bar\gamma\over\bar\gamma^{\beta_s}}\;
\int_{-1}^{1}{d\cos\alpha\over 2}\;
\delta[\gamma_{_A}-\bar\gamma\,\gamma\,(1+\cos\alpha)]\nonumber\\ && =
\int_{{\rm Max}\left[\gamma,\sqrt{\gamma_{_A}/(2\gamma)}\right]}^{b\gamma}\,
{d\bar\gamma\over\bar\gamma^{\beta_s}}\,
{1\over 2\,\gamma\,\bar\gamma}\; .
\label{AcceleratedCR}
\end{eqnarray}
This spectrum must still be integrated over the CB's trajectory, as in Eq.~(\ref{traj}),
and corrected for confinement in the Galaxy. The result is again simple and analytical,
but a bit long to report here. Below the knee, it has the same power-law behaviour as
the elastic contribution. The effect of the discontinuity in the lower limit of integration
in Eq.~(\ref{AcceleratedCR}) survives in the trajectory-integrated result as a predicted
smooth change in slope by $\Delta\beta\sim 0.3$ at $\gamma_{_A}=2\,\gamma_0^2$, 
which is what is observed, see Figs.~\ref{allpart},\ref{ffff}. 
The spectra extend all the way to the Larmor cutoff(s) of Eq.~(\ref{Larmor}). 

\subsection{Galactic confinement: the ankle(s)}

The interpretation of the ankle(s) in the CB model is conventional:
they are the $Z$-dependent energies at which the Galaxy
and its magnetized halo no longer confine cosmic rays \cite{DP}.
For $B\sim 3$ $\mu$G, the position of the ankle(s) is at:
\begin{equation}
E[{\rm ankle}]\sim 3\times 10^9\,Z\,{\rm GeV.}
\label{ankle}
\end{equation}
Cannonballs deposit CRs along their trajectories, reaching the halo and
beyond. Galactic CRs above $E\!=\!E[{\rm ankle}]$ escape. Instead of a cutoff,
a change to a harder spectrum is seen, which must therefore be an
{\it extragalactic flux}.  I have assumed in Figs.~\ref{allpart},\ref{ffff} 
that the spectrum of CRs above the ankles
is the source spectrum, corresponding to a sharp
transition from $c\sim 0.5$ to $c=0$ in Eq.~(\ref{c}).

\subsection{GZK modulations} 

Cosmic rays having travelled in intergalactic space along straight or curved 
trajectories for sufficiently long times should be subject to rather
sharp energy-cutoffs: the  GZK effect \cite{GZK}. Such cutoffs
would act as  ``chinese-lady's shoes'' further constraining the
``toe-nail'' cutoffs we  discussed. Are these GZK cutoffs expected in the
CB model? It depends on Galactic ``accessibility''.

It is difficult to ascertain the probability that extragalactic CRs of energies
{\it below} the ankle penetrate the Galaxy, if only because in the CB-model
there is an exuding Galactic ``wind'' of CRs and their accompanying magnetic
fields. If the Galaxy is quite ``accessible'', a good fraction of the observed
lower-energy CRs would be extragalactic (not a dominant fraction, for otherwise
redshift effects would erase the sharp features of the spectrum). A large
extragalactic contribution implies long ``look-back'' times and, consequently,
potentially observable GZK modulations. A small contribution implies 
short look-back times, no GZK effects, but the possibility of observing
relatively well-located point sources in the Virgo-cluster ``neighbourhood''.
Thus, in a sense, this is a ``no-lose'' situation: some new effect ought to be
found at the highest  energies.

\subsection{CR abundances}

\begin{figure}
\centering
 \epsfig{file=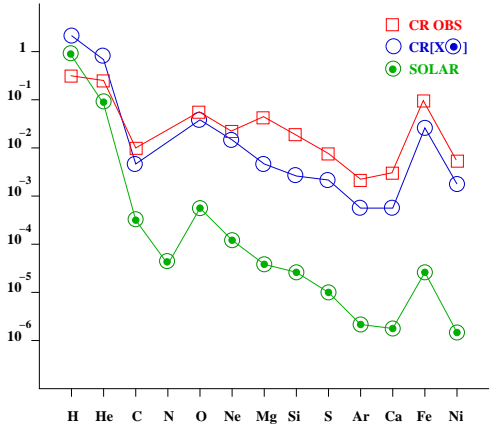,width=0.75\linewidth,angle=-90}
\vspace*{-16pt}
\caption{The relative abundances of primary CRs, from H to Ni.
The (green) dotted circles are solar-neighbourhood ISM abundances.
The (blue) circles are the predictions, with input solar abundances.
The (red) squares are observed CR abundances below $\sim 1$ TeV.}
\label{f1}
\end{figure}

Let $n_A$ be the number density of the ``target'' ISM nuclei converted
by the CBs' passage into CRs. The source spectra 
$dn_{_A}/d\gamma $,
are flavour-blind, so that the CR confinement-modified
energy spectra are of the form:
\begin{eqnarray}
{dN_{_A}\over dE}&
\propto&{1\over A\,m}\;{dn_{_A}\over d\gamma}\,\left({Z\over E}\right)^c
\nonumber\\
&\propto& K\,n_{_A}\, A^{\beta-c-1}\,Z^c\,E^{-\beta},
\label{abundances}
\end{eqnarray} 
with K a universal, composition-independent constant.
Below the knee(s) $\beta=2.7$ and $c\sim 0.5$ (I have ignored a
weak composition-dependence of $\beta$, discussed by Dar\cite{CRArnon}).
In Fig.~\ref{f1} the observed abundances 
of the most relevant {\it primary} CRs, up to Ni, are compared 
with the {\it solar} abundances, which are used as input to Eq.~(\ref{abundances}),
whose results are also shown.
In Figs.~\ref{allpart} and \ref{ffff}, I have referred to the
predicted and observed compositions of Fig.~\ref{f1} as ``light'' and ``heavy''.
The predictions are seen to fail at the large-metallicity end by a factor of $\sim 3$. 
This is what is expected, for CBs travel much of the time in a
star-burst region and a local superbubble that are known (within very
large errors) to have thrice the metallicity of the solar neighbourhood.
Given this uncertainty, it may be premature to do a complete calculation
taking into account CR propagation and
 the production of CRs with a broad spectral distribution
at the different points of many CB trajectories crossing a variety of ISM
domains. After all, our aim is to understand all of the salient features of the CR
conundrum, not its nitty-gritty details! 

\section{Conclusions and further predictions}

A large variety of high-energy astrophysical phenomena
are interrelated, and easy to understand in extremely simple terms \cite{Yo}. 
The unifying
concept is the ejection of relativistic blobs of matter in violent processes
of accretion onto compact objects. The inspiring observations are those
of quasars and microquasars. We have assumed that SNe 
axially eject very relativistic CBs of ordinary plasma, as their finite 
supply of matter  catastrophically accreting onto their central compact 
object is used up.
This assumption, complemented by a variety of observations of pre-SN
environments, explains not only the SN/GRB association, but the properties
of GRBs and XRFs \cite{GRB2}. The underlying process is ubiquitous in astrophysics:
(inverse) Compton scattering. None of the parameters involved in these 
predictions are put in by hand: they  rely on observations (e.g.
the early luminosity of a SN), or are borrowed from the CB-model analysis 
of GRB AGs (the distribution of CB Lorentz factors and of the CB-motion
angles relative to the line of sight to local observers, and the typical
mass of a CB).

The analysis of GRB AGs requires extra assumptions that are no doubt
over-simplifying: the way a relativistically expanding 
blob of plasma reaches an equilibrium radius as the process of radial
reemission of the ``collisionlessly'' scattered charged ISM particles
quenches the expansion, and the way in which the CB's magnetic-field 
pressure thereafter compensates the inwards force of the radially exuding 
particles. But the ensuing description of AGs as the synchrotron radiation
from the ISM electrons entering the CBs is simple and successful: the AG
light curves and wide-band spectra of all GRBs of known redshift are
well fit and, when predicted, correct.

Several predictions of the CB model of GRBs, we contend, are supported 
by the data, but require further corroboration. One  is the hyperluminal
motion of the CBs themselves \cite{GRB1,SL2}, which may be easier to detect in XRFs:
when not too far, they are simply GRBs seen at larger angles.
Another prediction concerns the AG X-ray lines, which in the
CB model are not the generally-assumed lines of Fe and other intermediate
elements, but the highly Doppler-boosted lines of light elements, notably
H Ly-$\alpha$ lines \cite{Xrays}. Since CBs decelerate in the ISM as they emit these
radiations, the lines should evolve towards lower frequencies in a
predicted fashion  \cite{Xrays}. 

On the basis of much less observational input, we propose \cite{GRB2} 
that short-duration
GRBs are associated with Type Ia SNe (30\% of the SNe are of this type,
30\% of GRBs are short).  If the observers did not give up so
early in attempting to discover the weak AG of short GRBs --but waited for
a few weeks for the peak SN light-- the SN ought to be observable. That would
be good news for cosmology, even if GRB-associated Type Ia SNe deviate
from the usual ``standard candle'' properties: in the CB model SNe are roughly
axially ---but not spherically--- symmetric. SN1998bw and the other almost
identical SNe associated with GRBs (some spectroscopically established),
are ordinary SNe seen very close to their axis. Both Type Ia and core-collapse
SNe (at least of Type Ib,c)
ought to be closer to standard ``torch-lights'' than to ``candles''.

We have also seen that the CB-model explains the shape of 
the CR electron spectrum, and the related spectrum and angular
distribution of the GBR, most of which is not ``cosmological'': it
 is associated at high latitudes with our own Galactic halo.
Higher-energy data on CR electrons and the GBR might confirm
the model by discovering the predicted ``knees'' in the corresponding
spectra \cite{GBR}. Seeing the ``GBR'' light from the halo of Andromeda would
also be quite a coup \cite{GBR}.

In the CB model, there are no Cooling Flows, but ``Warming Rays"
in dense X-ray emitting clusters.
It is the mysterious mechanism of {\it heating} that we  identified: 
CB-induced CRs \cite{CF}.

The CB model of CRs is rather
successful, considering that only one parameter was adjusted.
Clearly the results could be  improved, as better data
are gathered (e.g. on composition at all energies above $\!\sim\! 1\!$ TeV).
Many simplifying choices were made: a spatially constant
ISM composition, a 50-50 contribution of nuclei accelerated and unaccelerated
within a CB, a na\"ive energy dependence of the Galactic-confinement factor...
Even so, the distribution of CRs in the Galaxy, 
their total luminosity, the broken power-law spectra with 
their observed slopes, the position of the knee(s) and ankle(s), and the alleged 
variations of composition with energy are all explained in terms of 
simple physics. Surely, ``life'' may be more complicated, e.g.~nearby SNe
could contribute low-energy CRs, accelerated by conventional shock
mechanisms. There is no CB-model-specific
prediction concerning CRs, except that they are deposited
along very long lines exiting star-death regions, as opposed to points
in these regions, as in standard models. This prediction
might be testable in the search for line inhomogeneities
in the radiation from CR electrons.

The CB model is not a {\it theory} of practically all 
high-energy astrophysical phenomena.
It is lacking a deeper theoretical understanding of the magneto-dynamics within a
 CB; and of {\it cannons} themselves: the engines generating
the mighty ejections of compact astrophysical objects.

%\section*{Acknowledgements}
%
%I thank Shlomo Dado and Arnon Dar for a long and fruitful collaboration, and
%Nick Antoniou, Andy Cohen, Sergio Colafrancesco, Shelly Glashow and
%Rainer Plaga for collaboration and/or patient discussions.

\end{document}